\documentstyle[a4,aps,epsf]{revtex}
\def\uu{^{\mbox{ }}}
\def\s{\sigma}
\def\ss{\Sigma}
\def\up{\uparrow}
\def\dd{\downarrow}
\def\las{\langle}
\def\ras{\rangle}

\def\p{^{\prime}}
\def\kp{{\bf k_\parallel}}

\begin{document}
\title{Spin-polarized tunneling in ferromagnetic double barrier junctions}
\author{ Alireza Saffarzadeh\thanks{E-mail: a-saffarzadeh@cc.sbu.ac.
ir}\\ Department of Physics, Shahid Beheshti University, 19839,
Tehran, Iran}
\date{December 27, 2001}
\maketitle

\begin{abstract}
Spin-polarized tunneling in FMS/M/FMS double tunnel junctions where
FMSs are ferromagnetic semiconductor layers and M is a metal spacer is
studied theoretically within the single-site coherent potential
approximation (CPA). The exchange interaction between a conduction
electron and localized moment of the magnetic ion is treated in the
framework of the s-f model. The spin polarization in the FMS layers is
observed to oscillates as a function of the number of atomic planes in
the spacer layer. Amplitude of these oscillations decreases with
increasing the exchange interaction in FMS layers.
\end{abstract}

{\bf PACS.} 72.25.-b Spin polarized transport - 75.50.Pp Magnetic
semiconductors - 75.70.Ak Magnetic properties of monolayers and thin films

\section{\bf Introduction}
There is much recent interest in the spin-polarized transport in
multilayers of ferromagnets and paramagnets. These include the giant
magnetoresistance, spin-injection experiments, and spin-polarized
tunneling experiments which have application potential in digital
storage and magnetic sensor technologies \cite{Moodera1,Daughton}.
The spin-polarized tunneling phenomenon, showed that spin is conserved
in the tunneling process and the electrons coming from a ferromagnet are
spin-polarized \cite{Meservey}. In the last decade, with the progress in
the research on magnetic multilayers, spin-polarized tunneling
through magnetic semiconductor (MS) layers has
received increasing attention \cite{Metzke,Boru,Saffar}.
In tunneling experiments, when a FMS is
used as a tunnel barrier, the conduction band splits into up spin and
down spin subbands and the barrier height for these two subbands is
changed. Because of this exchange splitting, the probability of
tunneling for up spin electrons increases but for down spin it
decreases. Using MSs tunnel barriers such as EuS or EuSe, one can obtain
nearly 99\% spin-polarized tunneling electrons even with nonmagnetic
electrodes \cite{Moodera2,Moodera3}.

The purpose of this paper is to study the spin polarization of
the tunneling density of states in FMS double barrier junctions, and
show that it oscillates with the spacer thickness. The single-site CPA
for the s-f model in completely ferromagnetic case is used in the
calculations \cite{Takahashi1}. In order to determine the oscillations,
we estimate the difference in tunneling spin polarization between the
FMS single barrier and double barriers.

Although the CPA and the corresponding Alloy Analogy are not the best
starting points for treating the spin-polarization of conduction
electrons, we believe that the technique outlined in this article can
qualitatively recover the expected behavior for the spin polarization
as a function of the exchange coupling, the doping, and the spacer
thickness.

\section{\bf Model and formalism}
We consider a trilayer consisting of two FMS monolayers separated by a
nonmagnetic spacer. The trilayer is sandwiched between two semi-infinite
ideal lead wires as shown in figure 1. We assume that the interfaces
between the FMS layers
and the spacer are sharp. Both the trilayer and lead wires described by
a single-orbital tight-binding Hamiltonian with nearest-neighbor hopping
$t$ on a simple cubic lattice with lattice constant $a$. We choose
the (001) axis of the simple cubic structure to be normal to the layers
and this direction is called $z$-direction hereafter.

We use the s-f (or s-d) model which is commonly considered as realistic
for local-moment semiconductors and metals. In this model the following
Hamiltonian is used to describe the present system:
\begin{equation}
H=H_s+H_f+H_{sf}\  ,
\end{equation}
\begin{equation}
H_s=-t\sum_{{\bf r}n,{\bf r}\p n\p,\s}c^{\dag}_{{\bf r},n,\s}c_{{\bf r}
\p,n\p,\s}\uu\  ,
\end{equation}
\begin{equation}
H_f=-\sum_{{\bf r},{\bf r}\p,n}J_{{\bf r}n,{\bf r}\p n}
{\bf S}_{{\bf r},n}\cdot{\bf S}_{{\bf r}\p,n}\  ,
\end{equation}
\begin{equation}
H_{sf}=-I\sum_{{\bf r},n,\s,\s\p}
({\bf\mbox{\large$\s$}}\cdot{\bf S}_{{\bf r},n})_{\s\s\p}
c^{\dag}_{{\bf r},n,\s}c_{{\bf r},n,\s\p}\uu\  ,
\end{equation}
where $\bf r$ and $n$ denote the position in $x$-$y$ plane and the layer
index in the $z$-direction, respectively. Here $H_s$ is the transfer
energy of an s-electron with spin $\s (=\up ,\dd)$ between
nearest-neighbor sites. Each lattice point of the FMS layers is
occupied by a localized magnetic moment, represented by a spin operator
${\bf S}_{{\bf r},n}$. $H_f$ describes the Heisenberg-type exchange
interaction between these localized moments where
$J_{{\bf r}n,{\bf r}\p n}$ is an exchange integral. $H_{sf}$ is the s-f
exchange interaction between the s-electron and the f-spin
${\bf S}_{{\bf r},n}$ where $\bf\large\s$ is the Pauli matrix for the
conduction electron spin, and $I$ is the s-f exchange coupling constant.
In $H_{sf}$ and $H_f$, $n(=1, N)$ is the position of the FMS
layers in the $z$-direction. It is assumed that the localized moments of
the magnetic ions in two FMS layers to be the same magnetic moment. The
metal spacer is consisting of $N-2$ atomic layers which describes by
$H_s$.

In ordinary magnetic semiconductors, the magnetic excitation energy may
be smaller by two to three orders of magnitude than other typical
energies, as the Bloch bandwidth $W$ or the s-f exchange interaction
energy $IS$; thus, the f-spin is treated as a static system. On the other
hand, due to the Mermin-Wagner theorem \cite{Mermin,Gelfert},
an effectively two-dimensional spin-isotropic system
cannot display long-range magnetic order at finite temperatures,
$T>$ 0 K \cite{Schiller1}. This is one important reason why anisotropies
play a fundamental role for the understanding of thermodynamic phase
transitions in thin films. This restriction however, does not suppress
the main physical aspects at $T$=0 K. The spin splitting in density of
states of tunneling electrons is the main origin of the electron-spin
polarization, and is independent of the FMS layers thickness
\cite{Schiller2}. Thus we do not inspect how the spin system is affected
by the reduced translational symmetry.

The CPA \cite{Soven} was originally thought as an approximate theoretical
treatment of statistically disordered systems, e.g., binary alloys, but it
can easily be generalized to a random spin system if we ignore correlated
motion of localized spins \cite{Rangkubo}. In this investigation we use
the CPA for the s-f model in a single-site $t$-matrix formula, according
to Ref.\cite{Takahashi1}.

When an s-electron is propagating in the FMS layers it will subject to
different effective potentials through the s-f exchange interaction
according to the orientation of its spin. In order to treat the exchange
scattering of the s-electron within the framework of the single-site
CPA, we consider a single f-spin located at site {\bf r} in an effective
layered medium where an s-electron is subjected to a complex potential
(or coherent potential) which is site diagonal and takes the value
$\ss_{\up}$ or $\ss_{\dd}$, according to the spin orientation of the
s-electron \cite{Saffar}. Therein, an s-electron moving in this effective
medium can be described by the effective Hamiltonian $K$ in the
Bloch-Wannier representation as
\begin{equation}\label{k1}
K=\sum_{\kp,\s}\sum_{n,m}[(\ss_{n\s}+\epsilon_{\kp})
\delta_{n,m}-t(\delta_{m,n+1}+\delta_{m,n-1})]c^{\dag}_{\kp,n,\s}
c_{\kp,m,\s}\uu\  ,
\end{equation}
where $\ss_{n\s}$ is the layer- and spin-dependent coherent potential
which is only non-zero in the FMS layers. Here, $\kp(k_x,k_y)$ is a wave vector
parallel to the layers.

As in Refs. \cite{Saffar,Takahashi1}, we apply the condition that the average
scattering of the s-electron by the single f-spin in the medium is zero.
Thus we define the single-site $t$-matrix of the s-f exchange interaction as
\begin{equation}
t_{{\bf r},n}=v_{{\bf r},n}(1-\bar Gv_{{\bf r},n})^{-1}\  ,
\end{equation}
where $\bar G$ is the effective Green's function defined by
\begin{equation}\label{k2}
\bar G(Z)=\frac{1}{Z-K}\  .
\end{equation}
Here $t_{{\bf r},n}$
is the complete scattering associated with the isolated potential
$v_{{\bf r},n}$ in the $n$th effective layer ($n$=1 and $N$), which is
expressed as
\begin{equation}
v_{{\bf r},n}=\sum_{\s,\s\p}[-I({\bf\mbox{\large$\s$}}\cdot
{\bf S}_{{\bf r},n})_{\s\s\p}-\ss_{n\s}\delta_{\s\s\p}]
c^{\dag}_{{\bf r},n,\s}c_{{\bf r},n,\s\p}\uu\  .
\end{equation}

Within the single-site CPA, the condition
\begin{equation}
\las t_{{\bf r},n}\ras_{av}\uu=0  ,
\end{equation}
for any $\bf r$ in the FMS layers, leads to the equations for
$\ss_{n\up}(=\ss_\up$ in Ref.\cite{Takahashi1}) and $\ss_{n\dd}
(=\ss_\dd$ in Ref.\cite{Takahashi1}). Here the bracket
$\las\cdots\ras_{av}\uu$ means the thermal average.

In the completely ferromagnetic case (i.e. $T$=0 K) the orientations of the
f-spins are perfectly arranged in one direction ($z$-direction). In this
case the coherent potentials for two spin-polarized subbands are expressed
in the following simple forms
\cite{Takahashi1}:
\begin{equation}
\ss_{n\up}=-IS\  ,
\end{equation}
\begin{equation}
\ss_{n\dd}=IS\frac{(1+IF_{n\up})}{(1-IF_{n\up})}\  ,
\end{equation}
with
\begin{equation}\label{fn}
F_{n\s}(Z)=\frac{1}{N_\parallel}\sum_{\kp}\bar G_{nn\s}(\kp;Z)\  .
\end{equation}
Here, $\bar G_{nn\s}$ is the Green's function of the $n$th layer,
$N_\parallel$ is the number of lattice sites in each layer and
$Z=E+i\delta$, where $\delta$ is a small positive number.
Using the Eqs. (\ref{k1}) and (\ref{k2}), the Dyson equation in the
Bloch-Wannier representation can be written as
\begin{equation}\label{g1}
\bar G_{nm\s}=G_{nm\s}^0+G_{n1\s}^0\ss_{1\s}\bar G_{1m\s}+G_{nN\s}^0
\ss_{N\s}\bar G_{Nm\s}\  ,
\end{equation}
where
\begin{equation}
\bar G_{1m\s}=\frac{G_{1m\s}^0(1-G_{11\s}^0\ss_{N\s})+G_{1N\s}^0\ss_{N\s}
G_{Nm\s}^0}{(1-G_{11\s}^0\ss_{1\s})(1-G_{11\s}^0\ss_{N\s})-[G_{1N\s}^0]^2
\ss_{1\s}\ss_{N\s}}\  ,
\end{equation}
\begin{equation}\label{g2}
\bar G_{Nm\s}=\frac{G_{Nm\s}^0(1-G_{11\s}^0\ss_{1\s})+G_{N1\s}^0\ss_{1\s}
G_{1m\s}^0}{(1-G_{11\s}^0\ss_{1\s})(1-G_{11\s}^0\ss_{N\s})-[G_{1N\s}^0]^2
\ss_{1\s}\ss_{N\s}}\  ,
\end{equation}
and the unperturbed Green's function is given by
\begin{equation}
G_{nm\s}^0(\kp;Z)=\frac{1}{2t\sqrt{\eta^2-1}}[\eta-\sqrt{\eta^2-1}]
^{|n-m|}\  .
\end{equation}
Here,
\begin{equation}
\eta=(Z-\epsilon_{\kp})/2t\  ,
\end{equation}
\begin{equation}
\epsilon_{\kp}=-2t(\cos k_xa+\cos k_ya)\  .
\end{equation}
In Eqs. (\ref{g1})-(\ref{g2}), we have suppressed the variables
$\kp$ and $Z$ for simplicity. We have solved these equations numerically
for $\ss_{n\s}$. From equation (\ref{fn}) we can calculate the local density of
states (LDOS) per atomic site for spin $\s$ electron in the effective
layer $n$ as
\begin{equation}
D_{n\s}(E)=-\frac{1}{\pi}~\mbox{Im}~F_{n\s}(E +i\delta)  ,
\end{equation}
which should satisfy the following equation in all of the present
numerical calculations
\begin{equation}
\int_{-\infty}^{+\infty} D_{n\s}(E)dE=1.0\  .
\end{equation}

In order to study the tunneling spin polarization, we assume that
$N_\up/N_\dd$ is equal to $D_{n\up}(E)/D_{n\dd}(E)$, where
$N_\up(N_\dd)$ is the number of electrons with up (down) spin after
tunneling to the FMS conduction band, $D_{n\up}(E)(D_{n\dd}(E))$
is the LDOS with up (down) spin at the $n$th layer, and $E$
is a typical energy of the tunneling electrons. Thus the magnitude of
the spin polarization for tunneling density of states
in each layer can be given by \cite{Saffar,Takahashi2}
\begin{equation}
P_n=\frac{D_{n\up}(E_F)-D_{n\dd}(E_F)}{D_{n\up}(E_F)+D_{n\dd}(E_F)}\  ,
\end{equation}
where $E_F$ is the Fermi energy, because it is expected that only
electrons near the Fermi level participate in tunneling process.

We are mainly interested in the difference between the electron-spin
polarization in FMS/M/FMS and M/FMS/M junctions. Hence, in the
present results an effective polarization is used in place of $P_n$ which
is defined as
\begin{equation}
P_{eff}=P_{double}-P_{single}\  ,
\label{eff}
\end{equation}
where $P_{single}$ and $P_{double}$ are the layer dependence of spin
polarization ($P_n$) in M/FMS/M and FMS/M/FMS junctions respectively.
In fact we are interested in studying that part of the electron-spin
polarization which is due to the existence of the second FMS barrier (at
$n=N$); thus, it is convenient to cancel the contribution of M/FMS/M junction
to the spin polarization. In this way it is reasonable to discuss the effective
polarization. Note that in the M/FMS/M structure the FMS layer is at $n=1$.

\section{\bf Numerical results}
In the numerical calculations the energy is measured in units of $t$ and
the small imaginary part of the energy is chosen $\delta$=0.02 to
simplify the calculations.
The numerical results for the spin polarization in FMS layers as a
function of the spacer layer thickness $d=(N-2)a$ are shown in figure 2
for two cases: the localized moments in two FMS layers aline in
ferromagnetic (F) configuration ($\las S_{1}^{z}\ras_{av}\uu=\las S_{N}^{z}
\ras_{av}\uu=7/2$) and the moments aline in antiferromagnetic (AF)
configuration ($\las S_{1}^{z}\ras_{av}\uu=-\las S_{N}^{z}\ras_{av}\uu=7/2$).
These results are shown for various values $IS/W$ at $E_F=-5$. Here $IS/W$
which is the exchange-interaction strength, describes formally the strength
of the scattering processes in the ferromagnetic barriers.

This figure shows that the effective
spin polarization in FMS layers oscillates by increasing the spacer
thickness. The physical origin of such oscillations is attributed to
quantum interferences due to spin-dependent reflections of the electrons
at the FMS/M interfaces. The multiple interferences that take place in
the spacer, induce a change in the density of states of each subband.
Clearly, if the interferences in the spacer are constructive, one has an
increase of the density of states; conversely, when the interferences
are destructive, the density of states decreases. For the AF alignment,
where the magnetizations of the right
and left FMS layers are anti-parallel, electrons with up (down) spin are
easy (difficult) to tunnel into the spacer, and difficult (easy) to
tunnel out of it, because the densities of states of the left and right
FMS layers are different between up and down spin subbands (the inset of
figure 3 and 4). This imbalance among the tunnel currents causes the
spin accumulation, when the spin-relaxation time is sufficiently long
in the spacer. By increasing the $IS/W$, the band of up (down) spin is
shifted to the low (high) energy side and the splitting between these
subbands is increased \cite{Takahashi2}. In this case the $P_{single}$
is increased and the multiple interferences in the spacer and the
difference between the $P_{single}$ and the $P_{double}$ is reduced.
Thus, by fixing the $E_F$ for different values of $IS/W$, the amplitude
of oscillations of the effective spin polarization is decreased.

The effective spin polarization inside the spacer is shown
in figures 3 and 4 for the AF alignment. Here the spacer is consisting
of eight atomic layers. As the figures show, by decreasing the Fermi
energy, the effective spin polarization in the spacer decreases. It
confirms that because of the increase in spin splitting between up spin
and down spin subbands, the effect of multiple interferences in the
spacer and therefore the amplitude of oscillations are reduced.
Using these results, one can see that how long the
spin-polarized electrons remember their spin orientation. This is
especially important for electronic applications, because if the spins
relax too rapidly, the distances traversed by the spin-polarized current
in a device will be too short to serve any practical purpose.

\section{\bf Concluding remarks}
To summarize, on the basis of the single-site CPA for the s-f model
at $T$=0 K, we have investigated the spacer thickness dependence of
the tunneling spin polarization in FMS/M/FMS double tunnel junctions.
We have found that the spin polarization in the FMS layers oscillates
as a function of their separation. These oscillations is shown to
decreases with increasing the $IS/W$. This approach will be
improved by taking electron-magnon scattering into account, which
plays essential roles at low temperatures.
The present formulation is
applicable to problems of the interlayer exchange coupling and the
tunneling conductance in FMSs.

\newpage
\begin{figure}
\begin{center}
\leavevmode\hbox{\epsfxsize=1.1\textwidth\epsffile{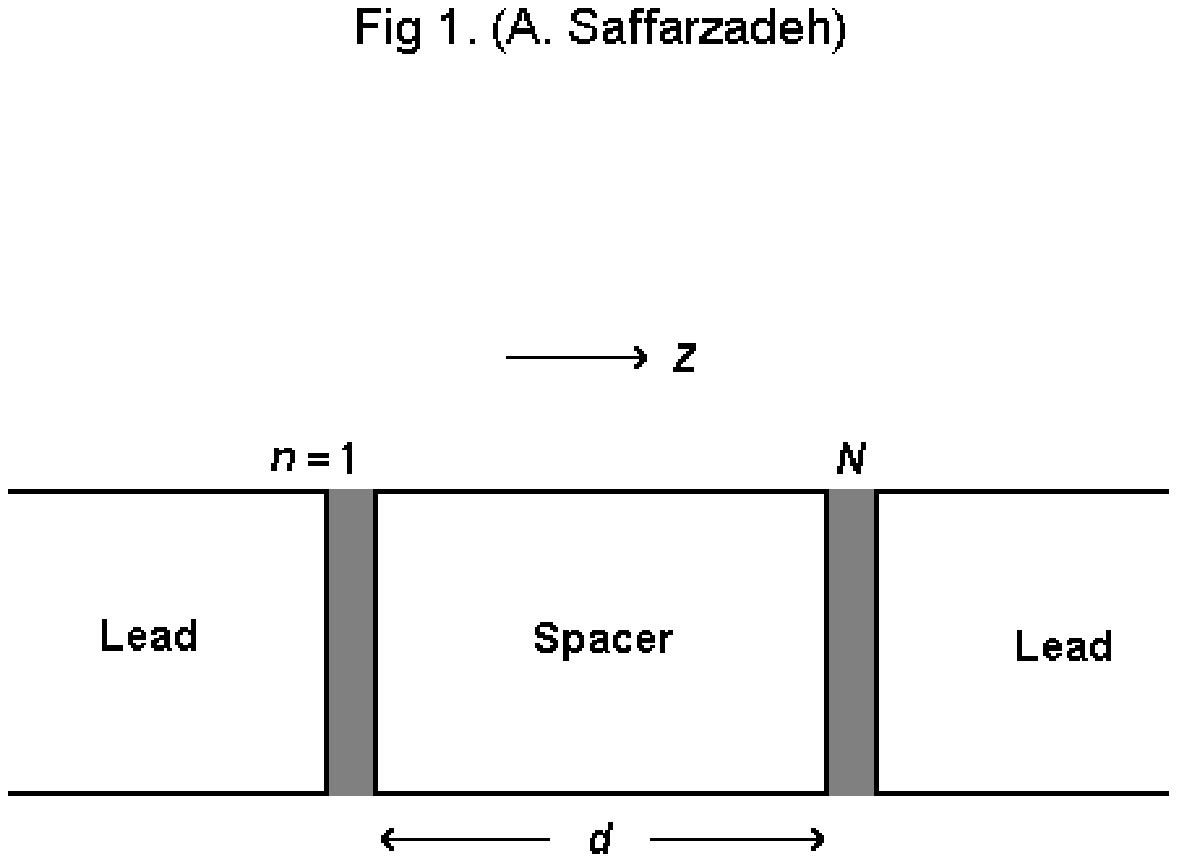}}
\end{center}
\caption{Schematic of the double junction. The atomic layers with
$n$=1 and $n$=$N$ are the FMS layers. Here $d$ denotes the spacer
thickness.}
\end{figure}
\newpage
\begin{figure}
\begin{center}
\leavevmode\hbox{\epsfxsize=1.1\textwidth\epsffile{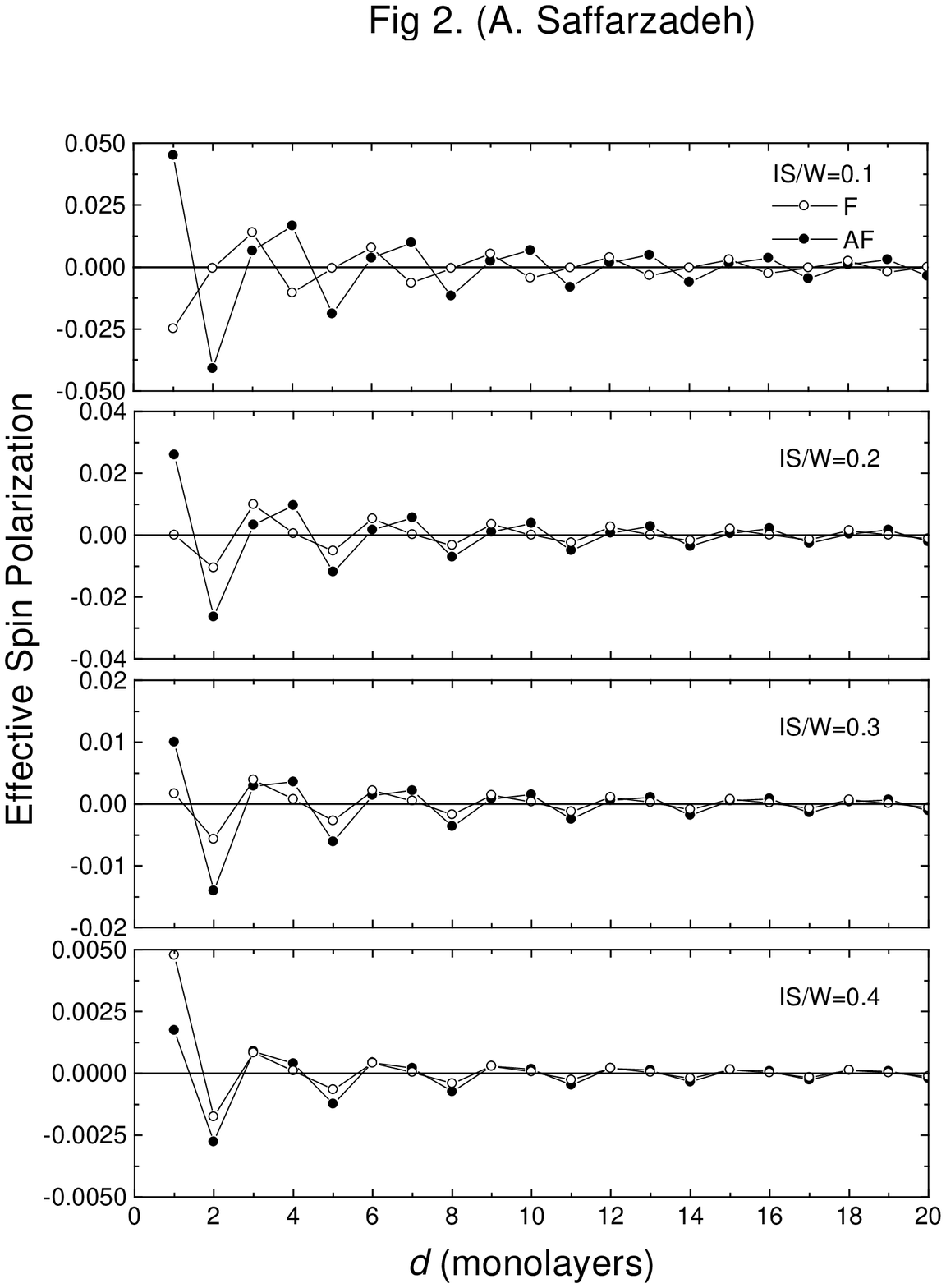}}
\end{center}
\caption{The effective spin polarization as a function of spacer
thickness $d$ in the AF and F alignments for $IS/W$=0.1, 0.2, 0.3, and
0.4. The Fermi energy is $E_F$=-5.0. Note the different scale on the
vertical axis.}
\end{figure}
\newpage
\begin{figure}
\begin{center}
\leavevmode\hbox{\epsfxsize=1.1\textwidth\epsffile{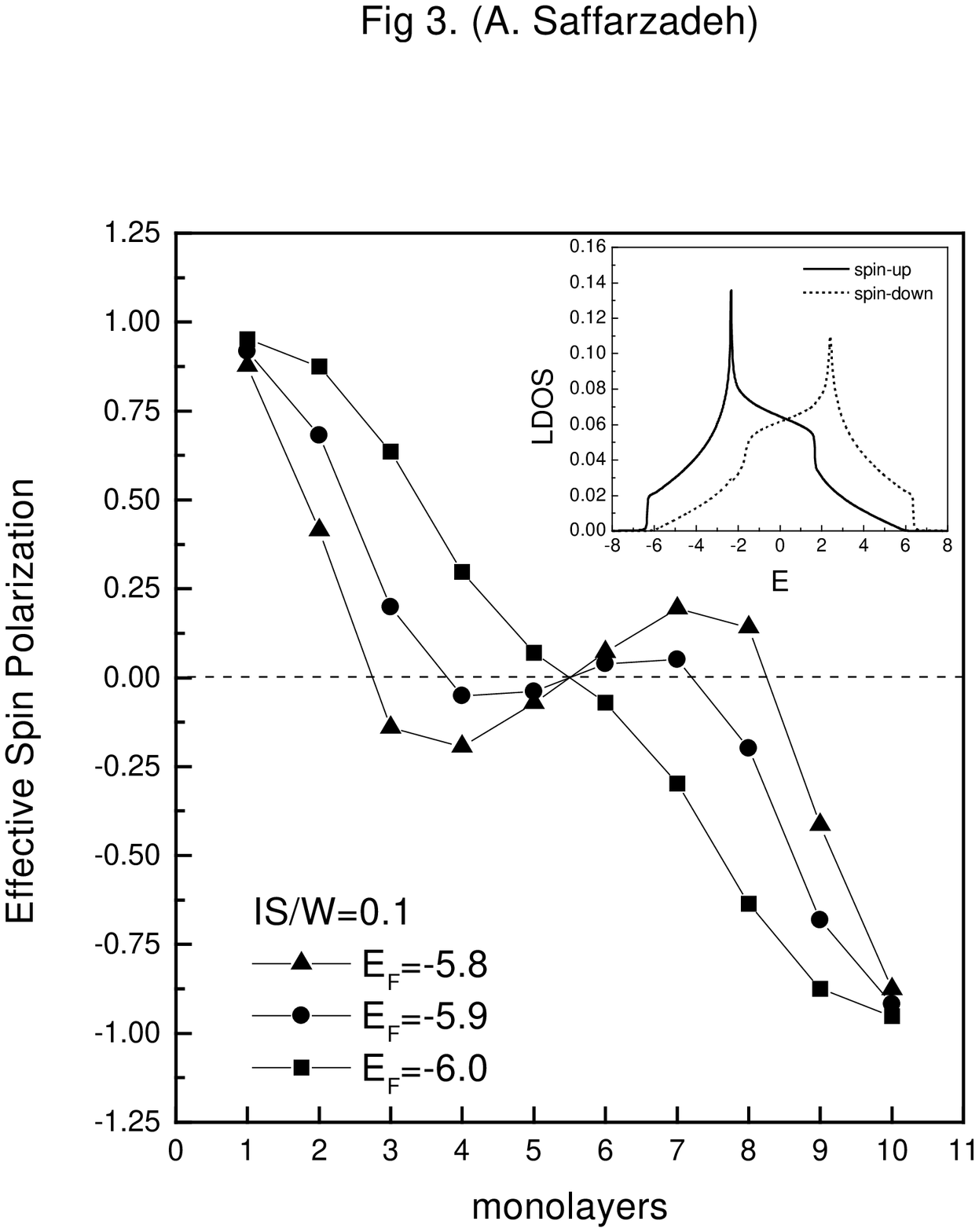}}
\end{center}
\caption{The effective spin polarization inside the spacer in the AF
alignment, with $IS/W$=0.1, $N$=10, and increasing values of $E_F$.
The inset is LDOS in FMS layers as a function of energy.}
\end{figure}
\newpage

\begin{figure}
\begin{center}
\leavevmode\hbox{\epsfxsize=1.1\textwidth\epsffile{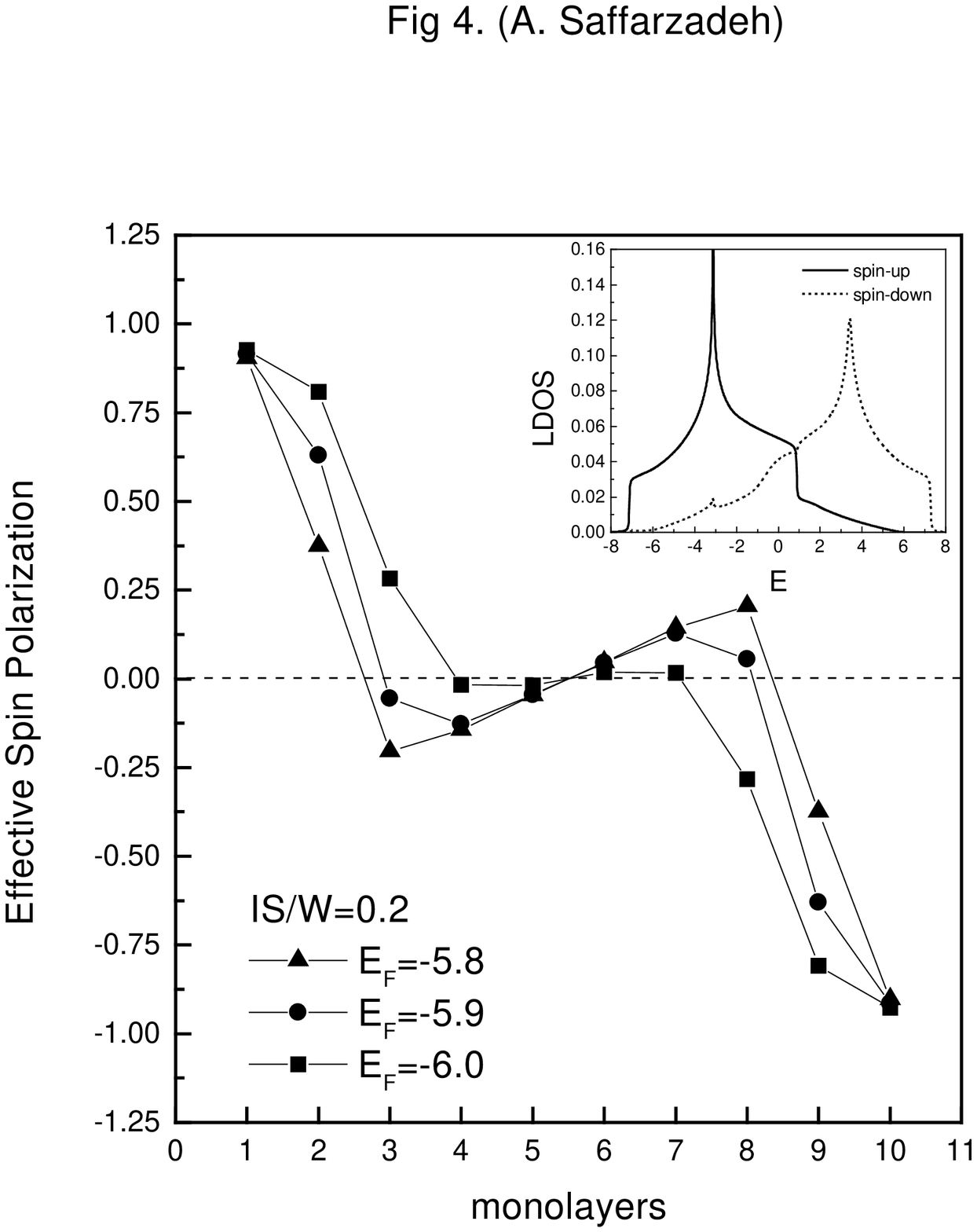}}
\end{center}
\caption{Same as in figure 3, but for $IS/W$=0.2\  .}
\end{figure}

\end{document}